\documentstyle[aaspp4,11pt]{article}

\def\gsim{\;\rlap{\lower 2.5pt
\hbox{$\sim$}}\raise 1.5pt\hbox{$>$}\;}
\def\lsim{\;\rlap{\lower 2.5pt
   \hbox{$\sim$}}\raise 1.5pt\hbox{$<$}\;}

\def\spose#1{\hbox to 0pt{#1\hss}}
\def\lta{\mathrel{\spose{\lower 3pt\hbox{$\mathchar''218$}}
     \raise 2.0pt\hbox{$\mathchar''13C$}}}
\def\gta{\mathrel{\spose{\lower 3pt\hbox{$\mathchar''218$}}
     \raise 2.0pt\hbox{$\mathchar''13E$}}}
\newcommand{\beq}{\begin{equation}}
\newcommand{\eeq}{\end{equation}}



\lefthead{MENOU, PERNA \& HERNQUIST} 
\righthead{HYDROGEN-POOR DISKS}

\begin{document}

\title{Hydrogen-Poor Disks in Compact X-Ray Binaries}

\author{Kristen Menou,\altaffilmark{1}}
 
\affil{Princeton University, Department of Astrophysical Sciences,
Princeton NJ 08544, USA, kristen@astro.princeton.edu}

\author{Rosalba Perna\altaffilmark{2,3} and Lars Hernquist}
 
\affil{Harvard-Smithsonian Center for Astrophysics, 60 Garden Street,
Cambridge MA 02138, USA, rperna@cfa.harvard.edu,
lhernquist@cfa.harvard.edu}
 
\altaffiltext{1}{Chandra Fellow}
\altaffiltext{2}{Harvard Junior Fellow} 
\altaffiltext{3}{Also at Istituto di Astrofisica Spaziale, C.N.R., 
via Fosso del Cavaliere, I-00133, Roma, Italy }

\begin{abstract}

We show that accretion disks in several compact X-ray binaries with
hydrogen--depleted donors are likely subject to a thermal ionization
instability, unless they are strongly irradiated. These disks are
particularly interesting in that their MHD-turbulent properties in the
neutral phase may be quite different from those of standard,
hydrogen--rich disks.

\end{abstract}

{\it subject headings}: X-ray: stars -- accretion, accretion disks --
MHD -- turbulence

\section{Introduction}

The vast majority of accretion disks around compact objects in close
binary star systems are composed of hydrogen--rich material, as a
result of Roche-lobe mass-transfer or wind-capture from a
non-degenerate companion star (see Lewin, van Paradijs \& van den
Heuvel 1995 and Warner 1995 for reviews). It has long been known that
such hydrogen-rich disks are subject to a thermal-viscous instability
occurring when hydrogen becomes partially ionized, at
central (mid-plane)
temperatures $T_c \sim 10^4$~K (or, equivalently, disk effective
temperatures $T_{\rm eff} \lsim 10^4$~K; Meyer \& Meyer-Hofmeister
1981; see, e.g., Cannizzo 1993a; Ludwig, Meyer-Hofmeister \& Ritter
1994; Hameury et al. 1998 for detailed calculations).

There is also a well-established class of close binaries, the
so-called AM CVn stars, in which the donor star is thought to be a
helium white dwarf (e.g. Warner 1995). Helium disks in these systems
are subject to the same thermal ionization instability as
hydrogen-rich disks, as shown by Smak (1983) and subsequent workers
(Cannizzo 1984; Tsugawa \& Osaki 1997; El-Khoury \& Wickramasinghe
2000).

Recently, the existence of a new class of close binaries, in which
disks are not made of hydrogen or helium but instead metals such as O
and Ne (because of the nature of the white dwarf donors), has been
proposed (Schulz et al. 2001; Juett, Psaltis \& Chakrabarty 2001). In
our study of the properties of supernova fallback disks (predominantly
made of metals; Menou, Perna \& Hernquist 2001a,b), we showed that disks
composed of pure C or O are also subject to the thermal ionization
instability. We argued that accretion disks of any metal content
should be subject to this instability as well.

In this {\it Letter}, we extend our work on metal disks to derive more
precise global stability criteria. We then apply these results to the
candidate metal disks in compact binaries, in an attempt to
characterize their stability properties.

\section{Local Thermal Structure}

The thermal-viscous stability of gaseous, thin accretion disks around
compact objects has been investigated extensively in the past.  We
follow a standard procedure here, similar to that employed to study
the properties of fallback disks (Menou et al. 2001a). Because of the
limited availability of metal opacities in the appropriate density and
temperature ranges, our work is limited to C and O compositions
only. The case of pure He is also considered for comparison with
previous results.

\begin{figure}
\plotone{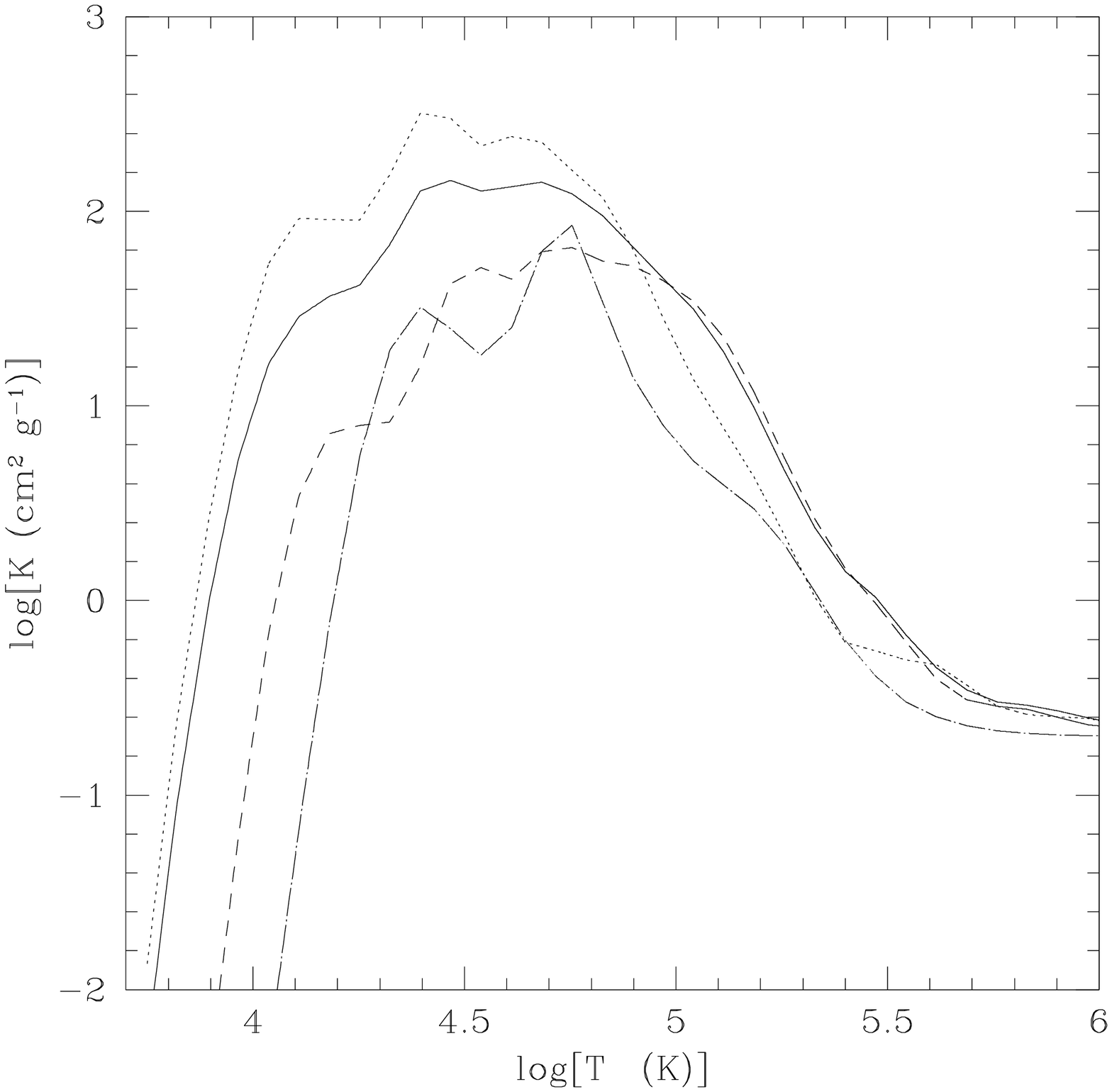}
\caption{Rosseland-mean opacities as a function of temperature, at a
mass density of $10^{-6}$~g~cm$^{-3}$, for pure Helium (dash-dotted),
pure Carbon (dotted), pure Oxygen (dashed) and a Carbon/Oxygen
composition with a 40\%-60\% mass fraction (solid). The sudden opacity
drop at $T \lsim 10^4$~K corresponds to the recombination of the last
available free electron. Note the similarity of the C/O opacity to the
Carbon one at low temperatures and to the Oxygen one at high
temperatures.
\label{fig:one}}
\end{figure}

The Rosseland-mean opacities are taken from the OPAL database
(Iglesias \& Rogers 1996)\footnote{The OPAL database website is {\tt
http://www-phys.llnl.gov/Research/OPAL/}}. The opacities, at a
specific mass density of $10^{-6}$~g~cm$^{-3}$, are shown as a
function of temperature in Fig.~\ref{fig:one}, for the four
compositions of interest: pure He (dash-dotted), pure C (dotted), pure
O (dashed) and a C/O composition with a 40\%--60\% mass fraction
(solid). The sudden opacity drop at temperatures $T \lsim 10^4$~K,
which corresponds to the recombination of the last free electron, is
responsible in each case for the thermal ionization instability (Menou
et al. 2001a). This temperature differs from composition to
composition because of different ground state ionization potentials
for each element. In particular, note that C dominates the opacity of
the C/O composition at low temperatures (while O does at high
temperatures). This is because C recombines at lower temperatures than
O, and it implies that the stability properties of a disk with this
C/O composition will be very similar to those of a disk made of pure C
(see below).

The disk thermal equilibria are found by equating the local viscous
dissipation rate $Q^+$ to the radiative cooling rate $Q^-$, where
$Q^-$ is a function of the disk opacity. We neglect X-ray irradiation
when calculating the thermal equilibria (see discussion below).  The
thermal equilibria are calculated from an extended grid of detailed
models for the disk vertical structure, for different values of the
surface density $\Sigma$ and the central temperature $T_c$, at a given
radius $R$ from the central object of mass $M_1$ (see Hameury et
al. 1998 for details on the numerical technique employed). Once a
solution to the vertical structure (corresponding to a given $\Sigma$,
$T_c$ and $R$) is known, the disk effective temperature $T_{\rm eff}$
is uniquely determined. The equation of state adopted for all the
compositions is that of a perfect gas with a monoatomic adiabatic
index $\gamma=5/3$. {The lack of a detailed equation of state
implies that vertical heat transport by convection, usually treated in
the mixing length approximation, cannot be included in our
calculations (see, e.g., Eq.~[22] of Hameury et al. 1998). We note,
however, that Gu, Vishniac \& Cannizzo (2000) have argued for weak
convection in thin accretion disks, because of the action of shearing
and turbulent radial mixing.}

\begin{figure}
\plotone{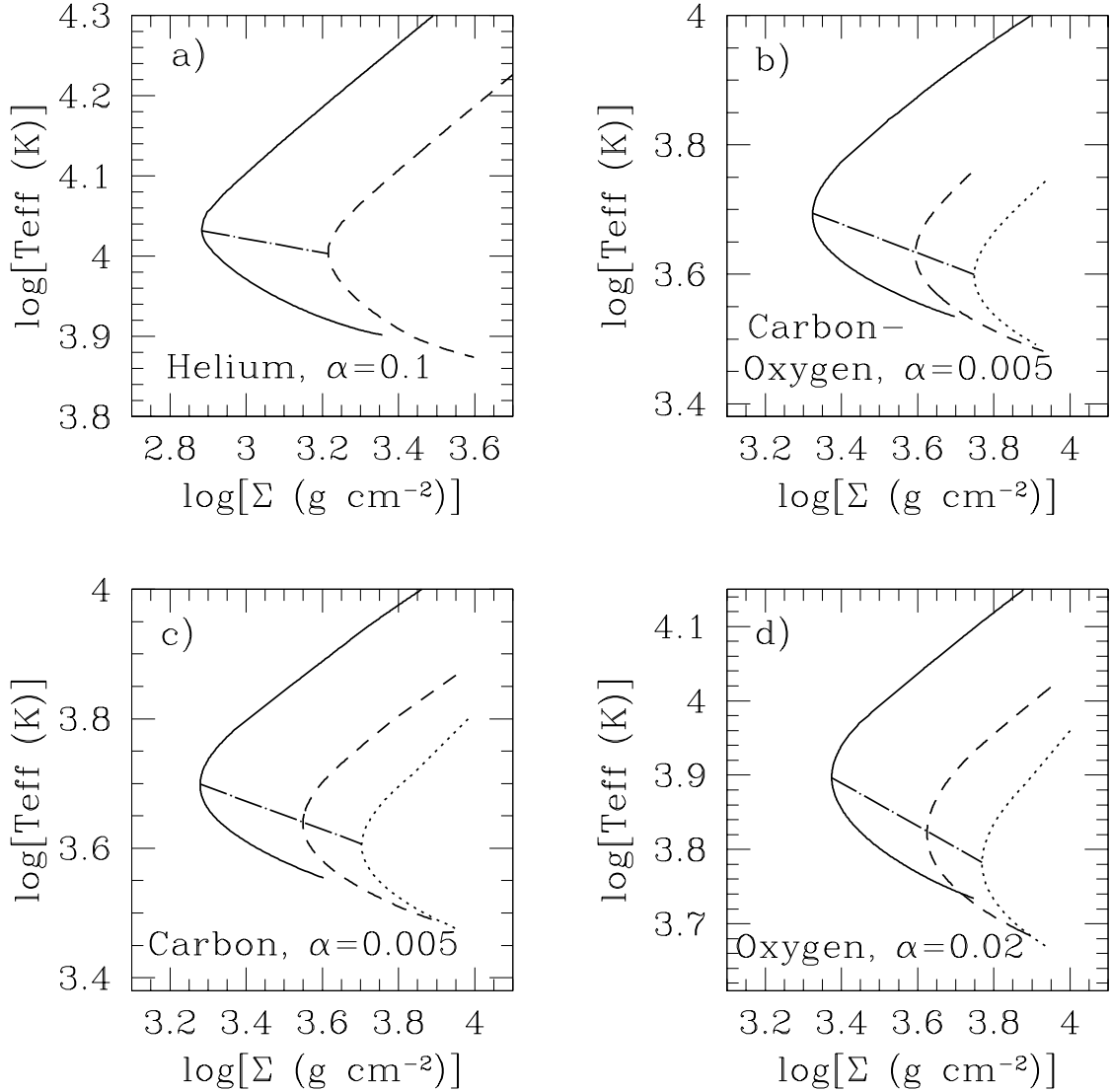}
\caption{Examples of thermal equilibrium curves (``S-curves'') for a
thin accretion disk, shown in a surface density vs. effective
temperature ($\Sigma - T_{\rm eff}$) diagram. The curves correspond to
a disk annulus located at $2 \times 10^{10}$~cm (solid line), $4
\times 10^{10}$~cm (dashed) or $6 \times 10^{10}$~cm (dotted) from a
central neutron star of mass $M_1=1.4 M_\odot$. The four panels
correspond to disks made of pure Helium (a), Carbon/Oxygen (40\%-60\%
mass fraction; b), pure Carbon (c) and pure Oxygen (d). The value
adopted for the viscosity parameter $\alpha$ and a fit to the critical
effective temperature are shown in each panel.
\label{fig:two}}
\end{figure}

Figure~\ref{fig:two} shows thermal equilibrium curves (``S-curves'')
thus obtained, in a surface density vs. effective temperature diagram.
In each panel, the sections of the S-curves with negative slope are
thermally and viscously unstable (see, e.g., Cannizzo 1993b and Lasota
2001 for reviews of the instability).  For each composition, the
thermal equilibrium solutions are shown for a different value of the
viscosity parameter $\alpha$, in order to guarantee that opacities are
available in the range of density and temperature covered by the
model. In the four panels of Fig.~\ref{fig:two}, the solid, dashed and
dotted lines correspond to ``S-curves'' calculated at a radius $R=2
\times 10^{10}$, $4 \times 10^{10}$ and $6 \times 10^{10}$~cm,
respectively, from a central object of mass $M_1 = 1.4
M_{\odot}$. (The calculation of only two S-curves for the He case was
sufficient to test the consistency with previous results.)

The instability occurs at a characteristic effective temperature
$T_{\rm eff, crit}$ (where the slope in the S-curve changes from
positive to negative), which is a function of the disk composition and
the radius of interest. When vertical transport of heat by convection
is not included, as is the case here (because of the simple equation
of state used), the value $T_{\rm eff, crit}$ is also function of the
viscosity parameter $\alpha$ (see below). As expected from the
C-dominated opacities at low temperatures, the S-curves for the C/O
composition are very similar to those of the pure C case. Our results
for the He case are in good agreement with previous work on He disks
(e.g. Smak 1983, Tsugawa \& Osaki 1997).

\section{Global Stability Properties}

The curves of thermal equilibrium can be used to derive a global
stability criterion for the corresponding accretion disk. The relation
$\dot M = 8 \pi R^3 \sigma T_{\rm eff}^4 / 3 G M_1$ between the disk
accretion rate and its effective temperature (e.g. Frank, King \&
Raine 1992) implies that a disk becomes thermally unstable if it
locally accretes at a rate below the value corresponding to $T_{\rm
eff, crit} (R)$.  The corresponding critical accretion rate is
therefore obtained by fitting the turnover point of the ``S-curve'' at
various radii for a given composition, as indicated by the dash-dotted
line in Fig.~\ref{fig:two}.

For the four compositions of interest here, we find { (after
extrapolation to large $\alpha$ for the three metallic cases)}:
\begin{eqnarray}
{\rm He:~~} \dot M_{\rm crit}^+ \left( R \right) & \simeq & 5.9
\times 10^{16} ~m_1^{-0.87} R_{10}^{2.62}~\alpha_{0.1}^{0.41}~{\rm
g~s}^{-1},\\
{\rm C:~~} \dot M_{\rm crit}^+ \left( R \right) & \simeq & 1.3
\times 10^{16} ~m_1^{-0.74} R_{10}^{2.23}~\alpha_{0.1}^{0.44}~{\rm
g~s}^{-1},\\
{\rm O:~~} \dot M_{\rm crit}^+ \left( R \right) & \simeq & 5.0
\times 10^{16} ~m_1^{-0.68} R_{10}^{2.05}~\alpha_{0.1}^{0.45}~{\rm
g~s}^{-1},\\
{\rm C/O:~~} \dot M_{\rm crit}^+ \left( R \right) & \simeq &
1.2 \times 10^{16} ~m_1^{-0.74}
R_{10}^{2.21}~\alpha_{0.1}^{0.42}~{\rm g~s}^{-1},
\label{eq:metalstab}
\end{eqnarray}
where $m_1$ is the mass of the central object in solar units, $R_{10}$
is the radius of interest in units of $10^{10}$~cm and {
$\alpha_{0.1}=\alpha / 0.1$}. The scaling of $\dot M_{\rm crit}^+$
with $m_1$ was deduced by noting that $m_1$ enters the problem only
via the square of the Keplerian angular frequency, $\Omega_K^2 \propto
m_1/R^3$, hence the direct proportionality to the scaling with radius
(see Hameury et al. 1998 for the equations solved).  We also have
included in the above formulae the results of numerical explorations
of the dependence of $\dot M_{\rm crit}^+ (R)$ on the viscosity
parameter $\alpha$ (not explicitly shown in Fig.~\ref{fig:two} for
conciseness).\footnote{We have checked that, e.g. in the
solar-composition case (for which a detailed equation of state was
available), the dependence of $\dot M_{\rm crit}^+$ on $\alpha$ nearly
disappears when significant vertical heat transport by convection is
allowed {(see Fig.~2 of Menou et al. 2001a for a specific,
quantitative comparison of stability in the presence and in the
absence of convection)}. This result, which is confirmed by several
previous vertical structure calculations (e.g. Cannizzo 1993b; Hameury
et al. 1998; Tsugawa \& Osaki 1997), shows that our poor understanding
of the role of convection for vertical heat transport in thin disks is
a significant source of uncertainty for the derived stability
criteria.}

These critical accretion rates do not differ strongly from the scaling
$\dot M_{\rm crit}^+ (R) \simeq 9.3 \times 10^{15} ~m_1^{-0.89}
R_{10}^{2.68}$~g~s$^{-1}$ found by Hameury et al. (1998) in the solar
composition case (for a viscosity parameter $\alpha=0.1$).  A disk of
a given composition which satisfies everywhere $\dot M(R) > \dot
M_{\rm crit}^+ (R)$ is ionized and thermally (as well as viscously)
stable.

\section{Applications}

In what follows, we apply the stability criteria derived in \S3 to a
few specific systems. In each case, we adopt a viscosity parameter
$\alpha=0.1$ since there is empirical evidence from dwarf novae
outbursts that this is the right order of magnitude for viscosity in a
fully-ionized, MHD-turbulent thin disk (Smak 1999; Gammie \& Menou
1998; Balbus \& Hawley 1991; 1998).

\subsection{Metal-Rich Disk in 4U 1626-67}

Schulz et al. (2001) recently reported the {\it Chandra} detection of
broad, double-peaked emission lines and strong photoelectric
absorption edges of O and Ne in the low-mass X-ray binary pulsar
4U~1626-67. Based on this, the authors argued that the disk in this
ultra-compact system (with $P_{\rm orb} = 42$~min; Middleditch et
al. 1981; Chakrabarty 1998) is probably fed by a C-O-Ne or O-Ne-Mg
white dwarf of mass $0.02 M_\odot$. The short orbital period and small
mass ratio in 4U~1626-67 imply a disk outer radius $\sim 2 \times
10^{10}$~cm (Schulz et al. 2001; Wang \& Chakrabarty 2001).

Assuming a mass $M_1 = 1.4 M_\odot$ for the neutron star and a
viscosity parameter $\alpha=0.1$, we find that the mass-transfer rates
below which the disk outermost regions in 4U~1626-67 become subject to
the thermal ionization instability are $\simeq 4.7 \times
10^{16}$~g~s$^{-1}$ and $\simeq 1.6 \times 10^{17}$~g~s$^{-1}$ for
pure C and pure O compositions, respectively. We note that the various
compositions discussed by Schulz et al. (2001) for the white dwarf
donor are dominated by either C or O, with some contribution from
Ne. The above estimates of critical mass-transfer rates for stability
should therefore be relevant in most cases because the disk stability
properties tend to be dominated by the element with the lowest
recombination temperature, as shown in \S2 and 3 (Ne recombines first,
followed by O and then C at lower temperatures; see e.g. Mihalas et
al. 1990).  However, if Mg makes a significant contribution to the
disk composition (a possibility suggested by Schulz et al. for an
O-Ne-Mg white dwarf donor), the critical mass transfer rate could be
even lower because Mg recombines at lower temperatures than C (we were
unable to address the Mg case in detail because opacities were not
available).

The above critical mass-transfer rates are between a factor a few and
two orders of magnitude larger than the observationally-inferred value
of $\sim 10^{16}$~g~s$^{-1}$ and the theoretically-inferred value of
$\sim 2 \times 10^{15}$~g~s$^{-1}$ discussed by Schultz et al. (2001)
for 4U~1626-67. Our work therefore suggests that the accretion disk in
this system should be subject to the thermal ionization
instability.\footnote{A similar conclusion is reached if the disk is
assumed to have a solar composition.}

\subsection{Hydrogen-Poor Disks in Other Systems}

Juett et al. (2001) propose that four other low-mass X-ray binaries
(4U~0614+091, 2S~0918-549, 4U~1543-624, 4U~1850-087) are also
ultra-compact binaries with low-mass, hydrogen-depleted donors. Only
4U~1850-087 has a known orbital period of $20.6$~min (clearly making
it an ultra-compact system), but Juett et al. argue that the three
other systems may also have orbital periods smaller than 1~hr.

If this interpretation is correct (i.e. short orbital periods, low
mass donors, helium- or metal-rich disks), we expect mass transfer
driven by gravitational radiation at a rate similarily low as in
4U~1626-67 and disks equally subject to the thermal ionization
instability.\footnote{Again, the presence of a substantial amount of
Mg in these disks could modify this conclusion.}  We note, however,
that the knowledge of the binary orbital period is essential for the
stability analysis because it largely determines the disk outer
radius, which enters the stability criteria in Eqs.~(1)--(4) with a
large power.

\subsection{Helium-Rich Disk in 4U 1820-30}

The low-mass X-ray binary 4U~1820-30 is an ultra-compact system, with
a $685$~s orbital period, located in the globular cluster NGC 6624
(Stella, White \& Priedhorsky 1987). The X-ray burst properties of
this system show that the donor star is a He white dwarf (Bildsten
1995). Given the short orbital period, the disk outer radius is $\lsim
10^{10}$~cm and, according to Eq.~(1), the critical mass-transfer rate
below which the He disk would be subject to the thermal ionization
instability is $\lsim 4.4 \times 10^{16}$~g~s$^{-1}$. The inferred
X-ray luminosity of the system is $> 2 \times 10^{37}$~erg~s$^{-1}$,
however, implying an accretion rate in excess of $10^{17}$~g~s$^{-1}$
(Bildsten 1995) and stability. This conclusion appears consistent with
the inferred stability of the accretion rate in the system (Bildsten
1995), although the $\sim 176$~d low-amplitude luminosity cycle
remains to be explained (Priedhorsky \& Terrell 1984).

\section{Discussion}

We have calculated, as a function of radius, critical accretion rates
below which disks become subject to the thermal ionization
instability, for four specific hydrogen-poor compositions. We have
applied these results to disks in several low-mass X-ray binaries with
possibly Hydrogen-depleted donors and found that some of them could
indeed be unstable.

We caution, however, that our results are subject to a number of
uncertainties. We have already mentioned the uncertainties associated
with the magnitude of convection in thin disks (our results correspond
to the inefficient convection limit). We have also shown that the
exact composition of the disk material is important because the
stability properties are in general dominated by the element with the
lowest ground state ionization potential (e.g. Mg for an O-Ne-Mg white
dwarf). Finally, we have neglected the stabilizing effects of disk
irradiation in our calculations (van Paradijs 1996; Dubus et
al. 1999). This is not unreasonable given that all the systems
discussed are thought to be ultra-compact, but we note that
irradiation may affect the stability of disks extending beyond
$10^9-10^{10}$~cm from the central neutron star, depending on their
geometry (e.g. Dubus et al. 1999).

The usual outcome of the thermal ionization instability in low-mass
X-ray binaries and dwarf novae is the development of large amplitude
outbursts and subsequent long periods of quiescence. However, this
outcome crucially depends on the efficiency of angular momentum
transport (or equivalently the value of the viscosity parameter
$\alpha$) in the disk when it becomes neutral (in the sense that a
value of $\alpha$ significantly smaller in the neutral disk than in
the ionized disk is required to produce large amplitude outbursts;
Smak 1984).\footnote{Several global numerical simulations have shown
that, for similar values of $\alpha$ in the neutral and ionized
phases, unstable disks experience only short, small amplitude
luminosity variations (Smak 1984; Menou, Hameury \& Stehle 1999).}
While there are good reasons for $\alpha$ to be significantly reduced
in a neutral, hydrogen-rich disk because of MHD turbulence decay
(Gammie \& Menou 1998; Menou 2000; Fleming, Stone \& Hawley 2000), it
may not be so in a hydrogen-poor disk.

Menou (2002) emphasizes the importance of dissociative recombination
by molecular hydrogen as a ``sink'' for free electrons and therefore a
``source'' of large resistivity in neutral, hydrogen-rich disks. In
disks lacking hydrogen, the decay of MHD turbulence may be reduced or
even absent, depending on the physical state and chemistry of the gas
in the neutral phase. In our view, this is one of the most interesting
aspects of the physics of hydrogen-poor disks, in the sense that a
differential study of the disk stability and variability properties in
hydrogen-poor and hydrogen-rich systems may reveal important clues on
the nature of viscosity in quiescent, neutral disks.

\section*{Acknowledgments} 

Support for this work was provided by NASA through Chandra Fellowship
grant PF9-10006 awarded by the Smithsonian Astrophysical Observatory
for NASA under contract NAS8-39073.  KM thanks the Center for
Astrophysical Sciences at Johns Hopkins University for hospitality.

\clearpage

\end{document}